\documentclass[12pt,a4paper]{article}
\usepackage{amsmath}
\usepackage{bm}
\usepackage{amsfonts}
\usepackage{graphicx}
\usepackage[normal]{caption2}
\usepackage{subfigure}
\usepackage{rotating}
\usepackage{citesort}
\usepackage{lscape}
\usepackage{section}
\begin{document}
\renewcommand\arraystretch{1.1}
\setlength{\abovecaptionskip}{0.1cm}
\setlength{\belowcaptionskip}{0.5cm}
\pagestyle{empty}
\newpage
\pagestyle{plain} \setcounter{page}{1} \setcounter{lofdepth}{2}
\begin{center} {\large\bf On the study of momentum correlations in fragmentation in heavy-ion collisions}\\
\vspace*{0.4cm}
{\bf Sakshi Gautam}$^{a}$\footnote{Email:~sakshigautm@gmail.com} {and \bf Rajni Kant}$^{b}$\\
$^{a}${\it  Department of Physics, Panjab University, Chandigarh
-160 014, India.\\} $^{b}${\it  House no. 276, Ward no. 11,
Tibbabasti, Patran, Distt. Patiala-147105, Punjab, India.\\}
\end{center}
The role of momentum correlations in fragmentation is studied
within the framework of quantum molecular dynamics model. Our
study is carried out by imposing momentum cut in the
clusterization algorithm. The study reveals a strong effect of
momentum cut in the fragmentation pattern. A comparison with
experimental data is also presented. Our theoretical results are
in agreement with the experimental data.
\newpage
\baselineskip 20pt
\section{Introduction}
 \par
  One of the most challenging questions in the present day
 nuclear physics research is the behavior of nuclear matter in a
 hot and dense region. This is not only important for nuclear
 physics, it is also useful for the understanding of the explosion
 mechanism of supernovae, the formation and structure of neutron
 stars. The heavy ion collisions at intermediate energies are
 excellent tool to study the nuclear matter at high density and
 temperature.
 \par
 At high excitation energies, the colliding nuclei may break up
 into several small and intermediate size fragments and a large
 number of nucleons are also emitted. This phenomenon is known as
 multifragmentation \cite{bege}. A large number of experimental attempts have
 been carried out ranging from the evaporation of particles to the
 total disassembly of the dense matter and a situation where
 excited matter breaks into several fragments.
The multifragmentation,is one of the rare phenomena that has
attracted major attention in recent years. The physics behind
multifragmentation is so complicated that many different
theoretical approaches have been developed
\cite{bege,aich1,qmd1,dorso}. Since no theoretical model simulates
fragments, one needs afterburners to identify clusters. Since
correlations and fluctuations are the main features of the
molecular dynamics model, the quantum molecular dynamics (QMD)
model is very successful in explaining the phenomena of
multifragmentation. Since every model simulate single nucleon, one
needs to have afterburner to clusterize the phase space. In a very
simple picture, we can define a cluster by using space
correlations. This method is known as minimum spanning tree (MST)
method \cite{jsingh}. In this method, we allow nucleons to form a
cluster if their centroids are less than 4 fm. This method works
fine when the system is very dilute. At the same time fragments
formed in MST method will be highly unstable (especially in
central collisions) as there the two nucleons may not be well
formed and therefore can be unstable that will decay after a
while. In order to filter out such unstable fragments, we impose
another cut in terms of relative momentum of nucleons. This
method, dubbed as minimum spanning tree with momentum cut (MSTP)
method was discussed by Puri \emph{et al.} \cite{kumar1}.
Unfortunately this study was restricted to heavier systems like
$^{93}$Nb+$^{93}$Nb and $^{197}$Au+$^{197}$Au reactions. The role
of momentum cut on the fragment structure of lighter systems is
still unclear. We aim to address this in present paper. Here we
plan to see the role of momentum cut on the fragment structure of
lighter colliding systems and also to see the role of colliding
geometry on the fragment
structure with momentum cut being imposed. \\
The present study is carried out within the framework of QMD model
\cite{aich1,qmd1} which is described in the following section.

\par
\section{The Formalism}
\subsection{Quantum Molecular dynamics (QMD) model}
\par
We describe the time evolution of a heavy-ion reaction within the
framework of Quantum Molecular Dynamics (QMD) model
\cite{aich1,qmd1} which is based on a molecular dynamics picture.
This model has been successful in explaining collective flow
\cite{sood2}, elliptic flow \cite{kumar3}, multifragmentation
\cite{dhawan} as well as dense and hot matter \cite{fuchs}. Here
each nucleon is represented by a coherent state of the form
\begin{equation}
\phi_{\alpha}(x_1,t)=\left({\frac {2}{L \pi}}\right)^{\frac
{3}{4}} e^{-(x_1-x_{\alpha }(t))^2}
e^{ip_{\alpha}(x_1-x_{\alpha})} e^{-\frac {i p_{\alpha}^2 t}{2m}}.
\label {e1}
\end{equation}
Thus, the wave function has two time dependent parameters
$x_{\alpha}$ and $p_{\alpha}$.  The total n-body wave function is
assumed to be a direct product of coherent states:
\begin{equation}
\phi=\phi_{\alpha}
(x_1,x_{\alpha},p_{\alpha},t)\phi_{\beta}(x_2,x_{\beta},
p_{\beta},t)....,         \label {e2}
\end{equation}
where antisymmetrization is neglected. One should, however, keep
in the mind that the Pauli principle, which is very important at
low incident energies, has been taken into account. The initial
values of the parameters are chosen in a way that the ensemble
($A_T$+$A_P$) nucleons give a proper density distribution as well
as a proper momentum distribution of the projectile and target
nuclei. The time evolution of the system is calculated using the
generalized variational principle. We start out from the action
\begin{equation}
S=\int_{t_1}^{t_2} {\cal {L}} [\phi,\phi^{*}] d\tau, \label {e3}
\end{equation}
with the Lagrange functional
\begin{equation}
{\cal {L}} =\left(\phi\left|i\hbar \frac
{d}{dt}-H\right|\phi\right), \label {e4}
\end{equation}
where the total time derivative includes the derivatives with
respect to the parameters. The time evolution is obtained by the
requirement that the action is stationary under the allowed
variation of the wave function
\begin{equation}
\delta S=\delta \int_{t_1}^{t_2} {\cal {L}} [\phi ,\phi^{*}] dt=0.
\label{e5}
\end{equation}
If the true solution of the Schr\"odinger equation is contained in
the restricted set of wave function
$\phi_{\alpha}\left({x_{1},x_{\alpha},p_{\alpha}}\right),$ this
variation of the action gives the exact solution of the
Schr\"odinger equation. If the parameter space is too restricted,
we obtain that wave function in the restricted parameter space
which comes close to the solution of the Schr\"odinger equation.
Performing the variation with the test wave function (2), we
obtain for each parameter $\lambda$ an Euler-Lagrange equation;
\begin{equation}
\frac{d}{dt} \frac{\partial {\cal {L}}}{\partial {\dot
{\lambda}}}-\frac{\partial \cal {L}} {\partial \lambda}=0.
\label{e6}
\end{equation}
For each coherent state and a Hamiltonian of the form, \\

$H=\sum_{\alpha}
\left[T_{\alpha}+{\frac{1}{2}}\sum_{\alpha\beta}V_{\alpha\beta}\right]$,
the Lagrangian and the Euler-Lagrange function can be easily
calculated
\begin{equation}
{\cal {L}} = \sum_{\alpha}{\dot {\bf x}_{\alpha}} {\bf
p}_{\alpha}-\sum_{\beta} \langle{V_{\alpha
\beta}}\rangle-\frac{3}{2Lm}, \label{e7}
\end{equation}
\begin{equation}
{\dot {\bf x}_{\alpha}}=\frac{{\bf
p}_\alpha}{m}+\nabla_{p_{\alpha}}\sum_{\beta} \langle{V_{\alpha
\beta}}\rangle, \label {e8}
\end{equation}
\begin{equation}
{\dot {\bf p}_{\alpha}}=-\nabla_{{\bf x}_{\alpha}}\sum_{\beta}
\langle{V_{\alpha \beta}}\rangle. \label {e9}
\end{equation}
Thus, the variational approach has reduced the n-body
Schr\"odinger equation to a set of 6n-different equations for the
parameters which can be solved numerically. If one inspects  the
formalism carefully, one finds that the interaction potential
which is actually the Br\"{u}ckner G-matrix can be divided into
two parts: (i) a real part and (ii) an imaginary part. The real
part of the potential acts like a potential whereas imaginary part
is proportional to the cross section.

In the present model, interaction potential comprises of the
following terms:
\begin{equation}
V_{\alpha\beta} = V_{loc}^{2} + V_{loc}^{3} + V_{Coul} + V_{Yuk}
 \label {e10}
 \end {equation}

$V_{loc}$ is the Skyrme force whereas $V_{Coul}$, $V_{Yuk}$ and
$V_{MDI}$ define, respectively, the Coulomb, and Yukawa
potentials. The Yukawa term separates the surface which also plays
the role in low energy processes like fusion and cluster
radioactivity \cite{puri}. The expectation value of these
potentials is calculated as
\begin{eqnarray}
V^2_{loc}& =& \int f_{\alpha} ({\bf p}_{\alpha}, {\bf r}_{\alpha},
t) f_{\beta}({\bf p}_{\beta}, {\bf r}_{\beta}, t)V_I ^{(2)}({\bf
r}_{\alpha}, {\bf r}_{\beta})
\nonumber\\
&  & \times {d^{3} {\bf r}_{\alpha} d^{3} {\bf r}_{\beta}
d^{3}{\bf p}_{\alpha}  d^{3}{\bf p}_{\beta},}
\end{eqnarray}
\begin{eqnarray}
V^3_{loc}& =& \int  f_{\alpha} ({\bf p}_{\alpha}, {\bf
r}_{\alpha}, t) f_{\beta}({\bf p}_{\beta}, {\bf r}_{\beta},t)
f_{\gamma} ({\bf p}_{\gamma}, {\bf r}_{\gamma}, t)
\nonumber\\
&  & \times  V_I^{(3)} ({\bf r}_{\alpha},{\bf r}_{\beta},{\bf
r}_{\gamma}) d^{3} {\bf r}_{\alpha} d^{3} {\bf r}_{\beta} d^{3}
{\bf r}_{\gamma}
\nonumber\\
&  & \times d^{3} {\bf p}_{\alpha}d^{3} {\bf p}_{\beta} d^{3} {\bf
p}_{\gamma}.
\end{eqnarray}
where $f_{\alpha}({\bf p}_{\alpha}, {\bf r}_{\alpha}, t)$ is the
Wigner density which corresponds to the wave functions (eq. 2). If
we deal with the local Skyrme force only, we get
{\begin{equation} V^{Skyrme} = \sum_{{\alpha}=1}^{A_T+A_P}
\left[\frac {A}{2} \sum_{{\beta}=1} \left(\frac
{\tilde{\rho}_{\alpha \beta}}{\rho_0}\right) + \frac
{B}{C+1}\sum_{{\beta}\ne {\alpha}} \left(\frac {\tilde
{\rho}_{\alpha \beta}} {\rho_0}\right)^C\right].
\end{equation}}

Here A, B and C are the Skyrme parameters which are defined
according to the ground state properties of a nucleus. Different
values of C lead to different equations of state. A larger value
of C (= 380 MeV) is often dubbed as stiff equation of state.The
finite range Yukawa ($V_{Yuk}$) and effective Coulomb potential
($V_{Coul}$) read as:
\begin{equation}
V_{Yuk} = \sum_{j, i\neq j} t_{3}
\frac{exp\{-|\textbf{r}_{\textbf{i}}-\textbf{r}_{\textbf{j}}|\}/\mu}{|\textbf{r}_{\textbf{i}}-\textbf{r}_{\textbf{j}}|/\mu},
\end{equation}
\begin{equation}
V_{Coul} = \sum_{j, i\neq
j}\frac{Z_{eff}^{2}e^{2}}{|\textbf{r}_{\textbf{i}}-\textbf{r}_{\textbf{j}}|}.
\end{equation}
\par
The Yukawa interaction (with $t_{3}$= -6.66 MeV and $\mu$ = 1.5
fm) is essential for the surface effects. The relativistic effect
does not play role in low incident energy of present interest
\cite{lehm}.
\par
\subsection{Minimum spanning tree (MST) method}
The phase space of nucleons is stored at several time steps. The
QMD model does not give any information about the fragments
observed at the final stage of the reaction. In order to construct
 the fragments, one needs
clusterization algorithms. We shall concentrate here on the MST
and MSTP methods.
\par
 According to MST method
\cite{jsingh}, two nucleons are allowed to share the same fragment
if their centroids are closer than a distance $r_{min}$,
\begin{equation}
|\textbf{r}_{\textbf{i}}-\textbf{r}_{\textbf{j}}| \leq r_{min}.
\end{equation}
where $\textbf{r}_{\textbf{i}}$ and $\textbf{r}_{\textbf{j}}$ are
the spatial positions of both nucleons and r$_{min}$ taken to be
4fm.
\par
\subsection{Minimum spanning tree with momentum cut (MSTP) method}
 For MSTP method,we impose a additional cut in the
momentum space, i.e., we allow only those nucleons to form a
fragment which in addition to equation(16) also satisfy
\begin{eqnarray}
|\textbf{p}_{\textbf{i}}-\textbf{p}_{\textbf{j}}| \leq p_{min},
\end{eqnarray}
where p$_{min}$ = 150 MeV/c.
\par
\section{Results and Discussion}
We simulated the reactions of $^{12}$C+$^{12}$C ,
$^{40}$Ca+$^{40}$Ca, $^{96}$Zr+$^{96}$Zr  and
$^{197}$Au+$^{197}$Au at 100 and 400 MeV/nucleon at $\hat{b}$ =
0.0, 0.2, 0.4, 0.6 and 0.8. We use a soft equation of state with
standard energy-dependent Cugon cross section.
\par
In Figure 1, we display the time evolution of A$^{max}$[(a),(b)],
free nucleons [(c),(d)] and LCPs(2$\leq$A$\leq$4) [(e),(f)] for
the reactions of $^{12}$C+$^{12}$C at 100 (left panels) and 400
(right) MeV/nucleon. Solid lines indicate the results of MST
method whereas dashed lines represent the results of MSTP method.
The heaviest fragment A$^{max}$ follows different time evolution
in MSTP as compared to MST method. In MST, we have a single big
fragment whereas momentum cut gives two distinct fragments which
shows realistic picture.
\par
In Figure 1(c) and 1(d), we display the time evolution of free
nucleons. We see that for both the energies, MSTP method yields
more free nucleons compared to MST method. There is also a delayed
emission of nucleons in MST because of no restrictions being
imposed. This delayed emission of free nucleons in MST method
taken place because of the fact that till 30 fm/c, we have a
single big fragment in MST method (See Figure 1(a),(b)). The
fragments saturate earlier in MSTP than MST as predicted in Ref.
\cite{kumar1}.
 \par
In figure 1 [(e),(f)], we display the time evolution of LCPs. We
see that MST yields more LCPs. The difference between MST and MSTP
method increases at 400 Mev/nucleon signifying significant role of
momentum correlations at higher incident energies.
\par
In figure 2, we display the time evolution of E$_{rat}$ of free
nucleons and LCP's for the reaction of $^{12}$C+$^{12}$C at
central (left panel) and peripheral (right) colliding geometry.
For MST and MSTP methods, we find a significant difference between
MST and MSTP methods for both free nucleons and LCP's. The
difference is more for the central collisions as compared to the
peripheral one.
\par
In figure 3, we display the impact parameter dependence of
A$^{max}$, free nucleons, LCPs, and IMFs for the reaction of
$^{40}$Ca+$^{40}$Ca at 100(left panel) and 400 (right)
MeV/nucleon. From both figures we see that A$^{max}$ rises with
impact parameter for both methods uniformly. The difference
increases with impact parameter. This happens because of the fact
that we have a bigger spectator matter (from where A$^{max}$
generates) at peripheral collisions geometry. The number of free
nucleons decreases with increase in impact parameter for both
methods.
\par
The emission of fragments shows that in contrast to central
collisions, peripheral collisions does not show drastic changes
with method. This happens due to fact that with increase in
colliding geometry, the fragments are the remnants of either
projectile or target, therefore breaking mechanisms are almost
bound and therefore MSTP methods does not give different results.
\par
In figure 4, we display the impact parameter dependence of
A$^{max}$, free nucleons, and LCPs for the reaction of
$^{12}$C+$^{12}$C at 100 (left panel) and 400 (right) MeV/nucleon.
We see that in this particular case, the effect of momentum cut on
the fragment production enhances with impact parameter (see figure
4(e) and (f)) which is quite different compared to earlier
figures. This is because the spectator matter even at peripheral
geometries will be very less in such a lighter system and so the
fragments are emitted mostly from the participant region, where
they are unstable and hence momentum cut plays a role at such
geometries.
\par
As a last step, we also compare our results with the experimental
data. In fig. 5, we display the charge distribution for
$^{197}$Au+$^{197}$Au reaction at 150 (left panels) and 250
(right) MeV/nucleon at central (b = 0-3.5) and semi-central (b =
0-8) colliding geometry. The data are taken from Refs.
\cite{sodan,wienold,kuhn}. Solid (open) circles represent the
results for MST (MSTP) method. From figure, we see that both MST
and MSTP methods obey the qualitative behaviour of charge
distribution, though, quantitatively, MST over predicts the data
at both the energies and at both colliding geometry. On the other
hand, we find that MSTP method matches the data well at both
energies and colliding geometries.

\section{Summary}
 Using the quantum molecular dynamic model, role of
 momentum correlations in fragmentation was studied. This was achieved by
 imposing cut in the momentum space during the process of clusterization. We find that this cut yields significant
 difference in the multifragmentation of colliding nuclei. Our
 findings are independent of the colliding energy of the reaction.

\section{Acknowledgement}
 This
work has been supported by a grant from Centre of Scientific and
Industrial Research (CSIR), Govt. of India.

\begin{figure}[!t]
\centering
 \vskip 1cm
\includegraphics[angle=0,width=12cm]{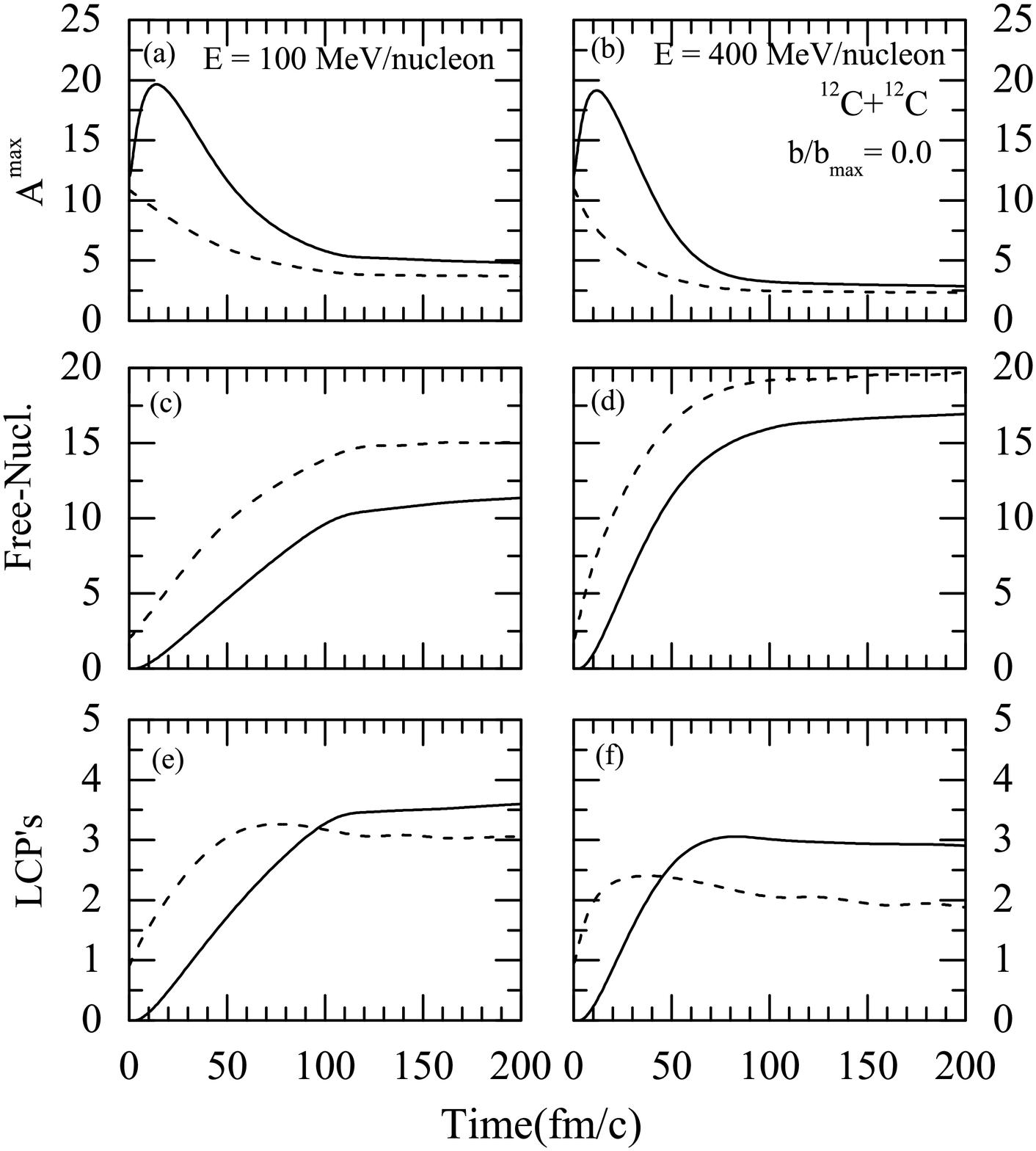}
 \vskip -0cm \caption{ The time evolution of A$^{max}$, free nucleons and LCPs for the reaction of $^{12}$C+$^{12}$C at incident energy of
 100 (left panels) and 400
MeV/nucleon (right) with MST and MSTP methods,
respectively.}\label{fig1}
\end{figure}

\begin{figure}[!t]
\centering
 \vskip 1cm
\includegraphics[angle=0,width=12cm]{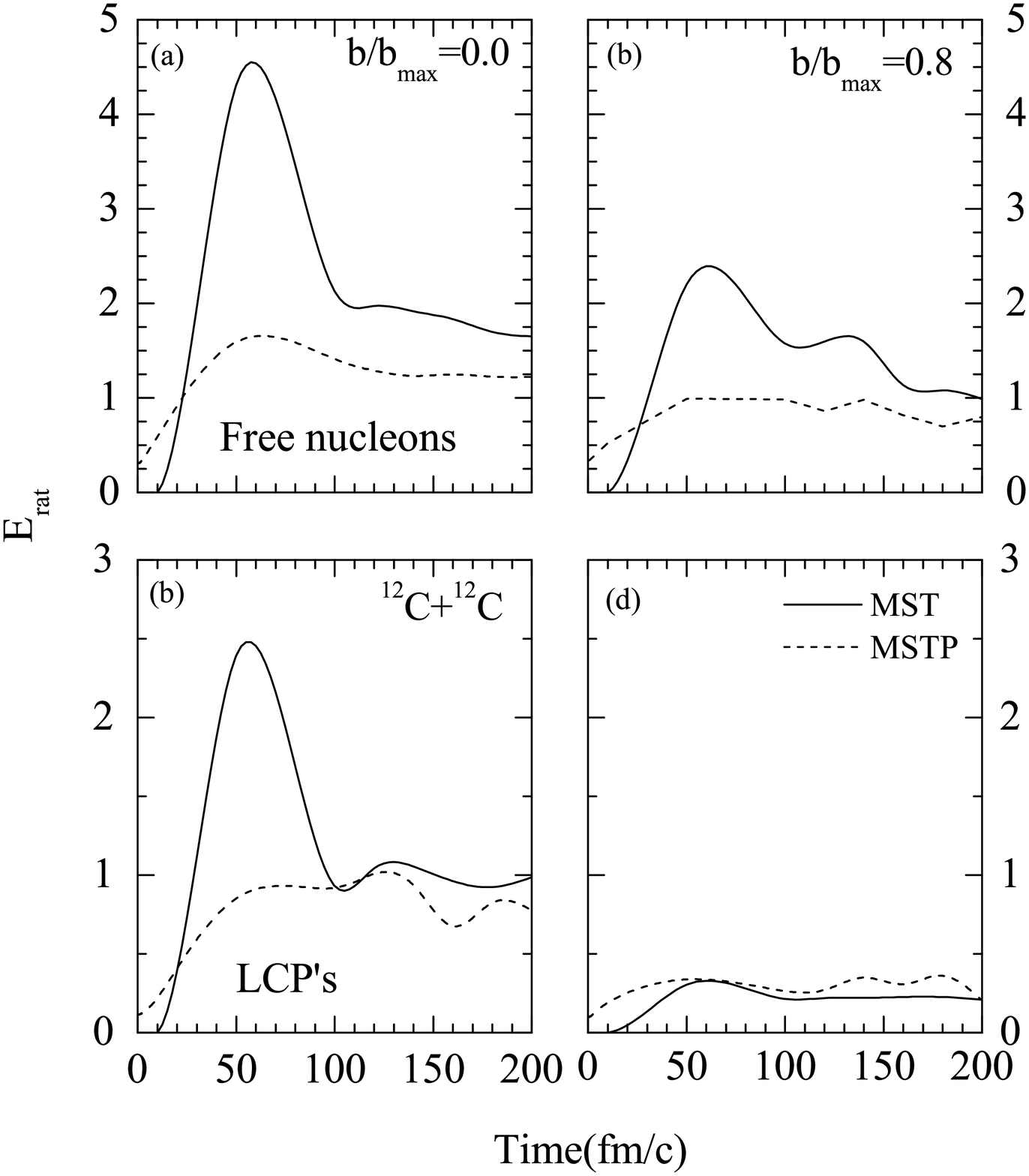}
 \vskip -0cm \caption{ The time evolution of E$_{rat}$ for free nucleons and LCPs for the reaction of
$^{12}$C+$^{12}$C at central (left panels) and peripheral (right)
collisions.}\label{fig6}
\end{figure}

 \begin{figure}[!t]
\centering \vskip 1cm
\includegraphics[angle=0,width=12cm]{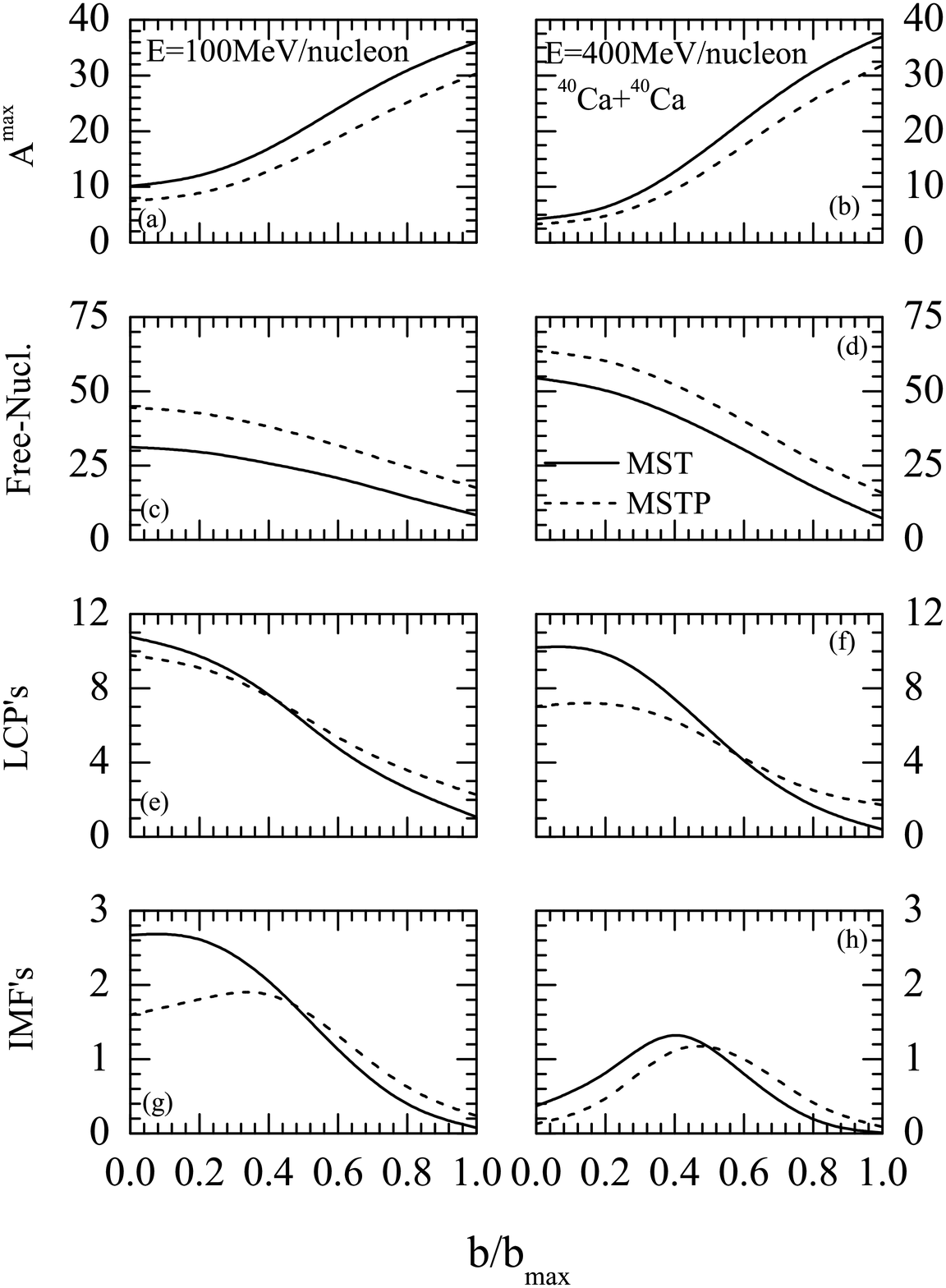}
\vskip -0cm \caption{The impact parameter dependence of A$^{max}$,
free nucleons, LCPs and IMFs for the reaction of
$^{40}$Ca+$^{40}$Ca at 100 (left panels) and 400 (right)
MeV/nucleon with MST and MSTP methods.}\label{fig3}
\end{figure}

\begin{figure}[!t]
\centering
 \vskip 1cm
\includegraphics[angle=0,width=12cm]{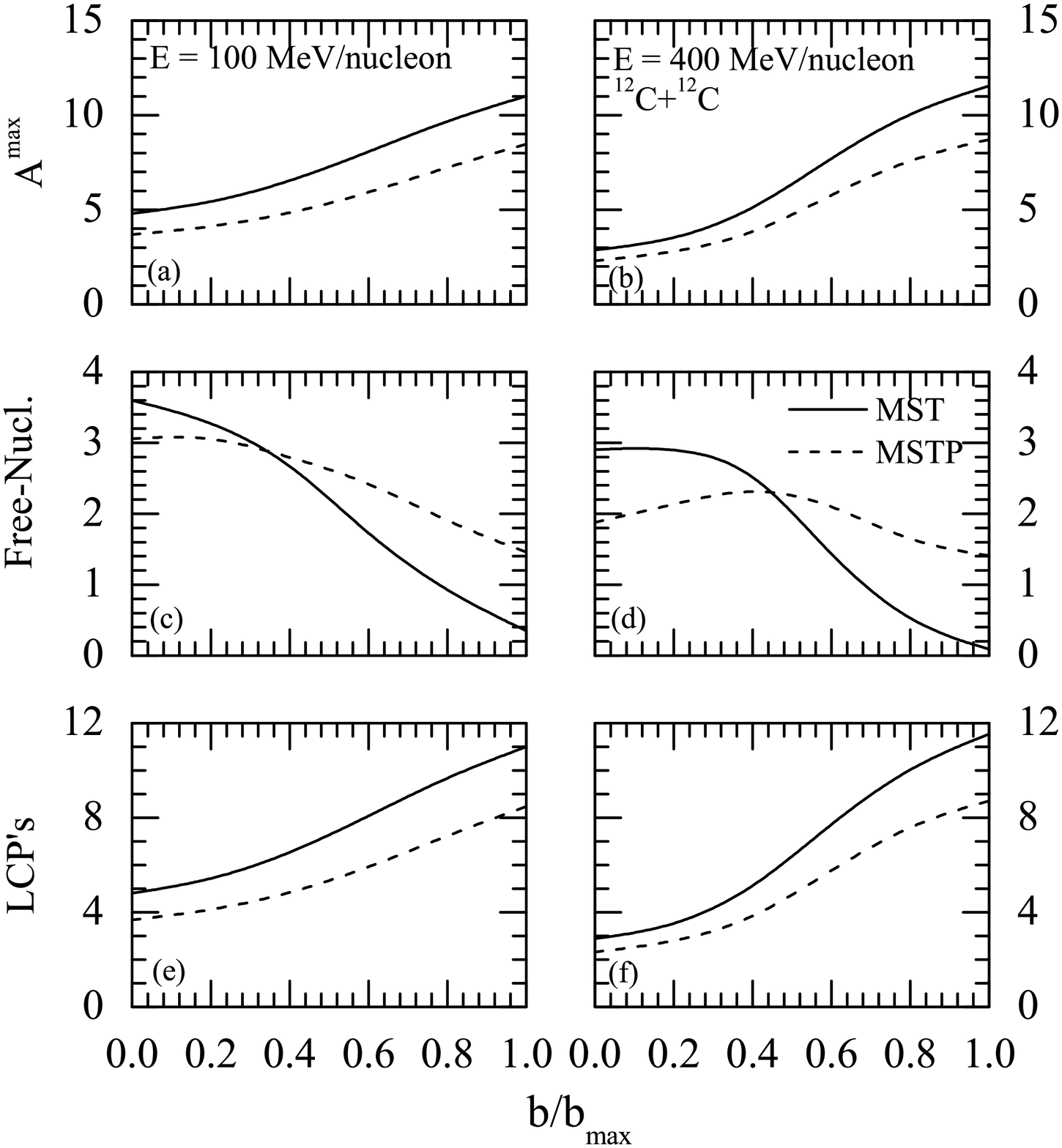}
 \vskip -0cm \caption{ Same as Fig. 3 but for the reaction of
$^{12}$C+$^{12}$C.}\label{fig6}
\end{figure}

\begin{figure}[!t]
\centering
 \vskip 1cm
\includegraphics[angle=0,width=12cm]{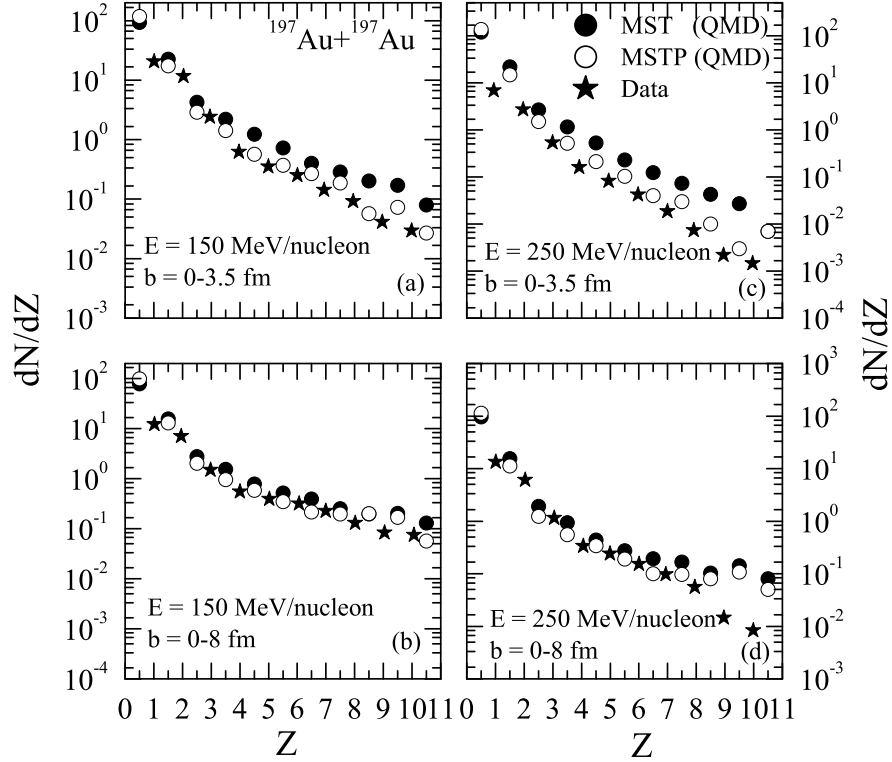}
 \vskip -0cm \caption{ Charge distribution for central and semi-central reactions of $^{197}$Au+$^{197}$Au reactions. The
 experimental values for central collisions at 150 MeV/nucleon are taken from Ref. \cite{kuhn} whereas the ones for semi-central
collisions at 150 MeV/nculeon are gathered from Ref.
\cite{wienold}. All experimental data for collisions at 250
MeV/nucleon are taken from Ref. \cite{sodan}. Circles represent
our theoretical calculations.}\label{fig6}
\end{figure}

\end{document}